\documentclass[english,prb,aps]{revtex4}
\usepackage{babel}
\usepackage{graphics}

\begin{document}
\title{Importance of Thermal Disorder on the Properties of Alloys:
Origin of Paramagnetism and Structural Anomalies in Iron-Aluminum}

\author{A. V. Smirnov and W. A. Shelton}

\affiliation{Computer Science and Mathematical Division,
Oak Ridge National Laboratory, Oak Ridge, TN 37831-6367}

\author{D. D. Johnson}

\affiliation{Materials Science and Engineering and Frederick Seitz
Materials Research Laboratory, University of Illinois, Urbana, IL
61801}

\date{January 28, 2004}

\begin{abstract}
The bcc-based Fe$_{1-x}$Al$_{x}$ exhibit interesting magnetic
and anomalous structural properties as a function of composition
and sample processing conditions arising from thermal or
off-stoichiometric chemical disorder, and, although well studied,
these properties are not understood.
In stoichiometric B2 FeAl, including the effects of partial long-range
order (i.e., thermal antisites), we find the observed paramagnetic
response (with non-zero local moments), in contrast to past investigations  
which find ferromagnetism based on local density
approximation (LDA) to density functional theory
or which find a  non-magnetic state from LDA+U,
both of which are inconsistent with experiment.
Moreover, from this magneto-chemical coupling, we are able to 
determine the origins of the  lattice constant anomalies found in 
Fe$_{1-x}$Al$_{x}$  for  x$\simeq 25-50$,
as observed from various processing conditions.
\end{abstract}

\maketitle

% Introduction
Although bcc-based iron-aluminum (Fe$_{1-x}$Al$_{x}$) alloys have been
investigated  extensively over the years, the complex (even anomalous)
structural and magnetic properties as a function of composition and state
of chemical order are not well understood.
In fact, while most experimental investigations observe Curie-Weiss-like
paramagnetic (PM) responses with small ($\sim0.3\mu _{B}$) effective
moments on Fe
\cite{BesnusHerr1975,ParthasarathiBeck1976,DomkeThomas1984,BognerSteiner1998},
electronic-structure calculations for perfectly ordered B2 ($\beta$-CuZn)  phase
obtain a ferromagnetic (FM) state with $\mu \sim 0.7\mu _{B}$
\cite{WilliamsKuebler1979,ChachamGalvao1987,SundararajanSahu1995,Bose1997,BognerSteiner1998,KulikovPostnikov1999}.
In addition, experimentally in the iron-rich Fe-Al system, anomalies are observed
in the average lattice constant that vary with processing conditions, i.e.,
whether the samples were quenched, annealed, and/or cold-worked
\cite{HumeRothery1969p172}.
The important fact is that the processing route used in  preparing samples
(stoichiometric or not)  produces lattice defects that  can have a significant
effect on the mechanical, structural and magnetic properties of Fe-Al as well as on
the intermetallics as a whole.  Furthermore, there are competing structures:
at $28\%~(38\%)$~Al at $825$~K ($475$~K), for example,  the B2 phase
undergoes a secondary-ordering transition to DO$_{3}$ (Fe$_{3}$Al or Heusler) phase.
And, due to kinetic limitations,  the phase diagram is known only for T $\geq 475$~K.
To be able to compare to experiment and to gain an understanding of the anomalies 
found in this system will require a first principles method capable of including disorder
and temperature effects on equal footing with the electronic structure.

{\par} Here, using first-principles methods, we investigate the structural 
and magnetic properties of bcc-based Fe-Al, including the A2 (bcc disordered), 
B2, and DO$_{3}$  phases, as a function of compositionally- and 
thermally-induced disorder (antisites).  
In B2 FeAl, we predict that the PM state competes with FM state
when the state of partial long-range order is taken into account. In particular,
at T$=0$ K, the PM and FM states are degenerate in energy at $\sim 70\%$
of perfect long-range order but are near-degenerate ($\leq 0.25$ mRy/atom)
over the range of $50-90\%$ of order (partial order compatible with
experiment); whereas FM state is $\sim0.5$ mRy/atom lower than the PM 
(and nonmagnetic (NM)) state for perfectly ordered B2, in agreement with 
past theoretical work.
More generally, this magneto-chemical coupling leads to
anomalies in the average lattice constant
in Fe$_{1-x}$Al$_{x}$ for  $x \simeq 0.25-0.5$ at.\%Al,
reproducing what has been observed for annealed, quenched and
cold-worked samples, but also never explained or found from theory.
Our results provide a simple and physically reasonable understanding of the,
heretofore, anomalous properties found in bcc-based Fe-Al.

\section{Discussion}
{\par}
Ordered intermetallics constitute an important class of high-temperature
structural materials. One of the most common crystals structures in the
intermetallics is the B2 structure, which because of its simple lattice
structure makes it an ideal model for studying a variety of physical
phenomena found in these systems \cite{BakerMunroe1990}.
The physical properties of Fe-Al are quite sensitive to extrinsic
defects \cite{LiuLee1989, TakahashiUmakoshi1991} and thermomechanical history
\cite{HumeRothery1969p172,NagpalBaker1990,ChangPike1993,WeberMeurtin1997}.
In fact, B2 FeAl is not fully ordered due to various lattice defects,
such as vacancies and antisites, and they are thought to play a major
role in the mechanical and magnetic behavior of this system.
However, the precise concentration of the different lattice defects is not
well known. Furthermore, unlike the strongly ordered intermetallics
such as B2 NiAl, which forms triple defect structures consisting of
two vacancies on the transition metal sublattice and an antisite defect
on the Al sublattice, the possible defect structures for the less
strongly ordered B2 FeAl are much more complicated, as the formation
of both vacancies and antisite defects on the Fe site are thermodynamically
stable\cite{FuYe1993}.

{\par} As mentioned,  using the local density approximation (LDA) to 
density functional theory (DFT) \cite{DFTReviewLundqvist1983}, 
electronic-structure calculations on ideal B2-FeAl obtain a FM state
with $\mu \sim 0.7\mu _{B}$ 
\cite{WilliamsKuebler1979,ChachamGalvao1987,SundararajanSahu1995,BognerSteiner1998,KulikovPostnikov1999}, 
in contradiction to the Curie-Weiss-like PM response found in experiment.
Calculations based on the so-called disordered local moment (DLM) 
paramagnetic state have also been suggested, but this approach by itself
was unable to explain experiment \cite{Bose1997}.
Calculations using generalized gradient approximation (GGA) to DFT,
which are known to improve the magnetic ground-state description
of elemental bcc Fe,\cite{ElsaesserZhu1998} have been employed but
do not yield FM DO$_{3}$-Fe$_{3}$Al or NM B2-FeAl  
as a ground-state \cite{LechermannWelsch2002}.
On the other hand, recent investigations \cite{MohnPersson2001,PetukhovMazin2003}
have obtained a NM ground-state for ideal B2 FeAl within so-called
LDA+U procedure, which is an extension of LDA that introduces
correlation corrections due to the orbital dependence of the Coulomb interaction.
The LDA+U finds a NM ground-state for B2-FeAl for a
particular range of  the empirically chosen parameter U.
However, once again, it is in contradiction to experiment, and does not confirm
the LDA+U picture.
Furthermore,  none of these investigations address any other experimental
observation, such as, the important  composition-dependent structural 
anomalies strongly affected by processing.

{\par} In alloys, both short-range order and partial long-range order (LRO),
and associated magnetism, are strongly affected by the processing temperatures,
whether its an  anneal or quench, which freezes in chemical disorder at
that temperature
(e.g. 0 $<$ T$_{Curie}$ $<$ $T_{processing}$ $<$ T$_{order-disorder}$).
For example, DO$_3$-Fe$_{3}$Al  crystal structure has $\sim 8\%$ site
disorder according to Bradley and Jay \cite{BradleyJay1932}, which was
as later confirmed \cite{TaylorJones1958,LawleyCahn1961}.
Thus, the overall processing procedure affects greatly the materials properties
and, therefore, the characterization data obtained from samples that are not
completely ordered (due to kinetic limitations).
The aforementioned DFT calculations provide results for perfectly
ordered states or off-stoichiometric disorder corresponding to T=0 K, which
may not be directly related to the experimentally assessed data.
However, Johnson et. al. \cite{JohnsonPRBRC2000} have shown that
including the thermally-induced partial LRO effects allows quantitative
comparisons to characterization experiments, and explains discrepancy of
$T=0$ K DFT results.
Clearly an investigation of the effects of (thermally-induced or off-stoichiometry)
partial order on the magnetic and structural properties in the B2 FeAl system
is in order.

\section{Computational Details}
{\par}
For our electronic-structure calculations, we use the Green's function based,
multiple-scattering approach of  Korringa, Kohn and Rostoker (KKR)
\cite{Korringa1947,KohnRostoker1954}.
We use the local density approximation to the exchange-correlation
potential and energy as parameterized by Von-Barth-Hedin
\cite{BarthJPhysC1972}.
In addition, all calculations are performed using the atomic sphere
approximation (ASA) with equal size Fe and Al atomic spheres
\cite{CalcDetails}.

{\par} By using the KKR method, we can include chemical and magnetic disorder
in the electronic structure and energetics via the 
coherent potential approximation\cite{Johnsonetal1986,Johnsonetal1990} (CPA)
modified to incorporate improved metallic screening due to
charge-correlations arising from the local chemical environment
\cite{JohnsonPinski1993}.
The KKR-CPA density-functional theory and total energy formalism
provides reliable and well-documented
energetics and structural-related parameters, from fully disordered to ordered
configurations. 

{\par} We describe the PM state by the  disorder local moment (DLM)
approximation\cite{Hubbard1983,Hasegawa1979,Staunton1985},
where magnetic short-range order is ignored, yet the site-dependent average
effect of magnetic orientational disorder is self-consistently included, whether
chemically ordered (partially or fully) or disordered.
Thus, our implementation of the KKR-CPA is capable of treating 
simultaneously chemical disorder, site defects and magnetic 
(orientational) disorder all on equal footing with the electronic structure 
(e.g., Slater-Pauling curve \cite{Johnson1987},
or the INVAR  \cite{Johnsonetal1989,JohnsonShelton1997}).
\begin{figure}
{\centering \resizebox*{1.5in}{2.2in}{\includegraphics{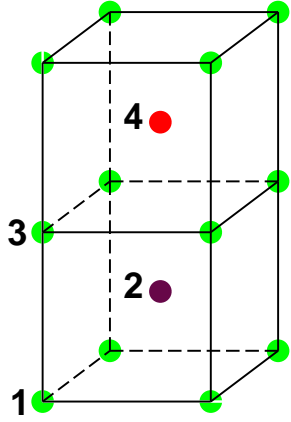}} \par}
%{\centering \resizebox*{1.5in}{2.2in}{\includegraphics{FeAlFig1.pdf}} \par}
\caption{The four-atom cell used to study cubic DO$_{3}$.
For basis atoms \emph{1-4}, the translational vectors
in lattice units of  T$ _{1}$=(101), T$_{2}$=(011), T$_{3}$=(001)
generate the standard 16-atom (cubic) DO$_{3}$ cell.
In ideal DO$_{3}$ Fe$_{3}$Al,  Fe (Al) occupies sites \emph{1-3}
(site \emph{4}); whereas in B2 FeAl, Fe (Al) atoms occupy odd (even)
numbered site. Sites \emph{1} and \emph{3} are always equivalent.}
\label{FigCell}
\end{figure}

{\par} For all calculations we use a four-atoms unit cell (see Fig. \ref{FigCell})
based on an  underlying bcc crystal lattice that inherently contains the 
DO$_{3}$, B2 and A2 (bcc disordered) structures  as a function of 
long-range order or composition because each sublattice 
is  composed locally of Fe$_{1-x}$Al$_{x}$.
Using Fig. \ref{FigCell} it is easy to define order parameters needed.
For example, referenced to the Al sublattice ($i=$Al), the B2 ordering at x=1/2
can be defined in terms of a temperature-dependent LRO parameter $\eta$(T):
$c_{i=Al}^{Al}=\frac{1}{2}(1+\eta(T))$ and
$c_{i=Al}^{Fe}=\frac{1}{2}(1-\eta (T))$, giving
the fully disordered A2 (ordered B2) lattice for $\eta$=$0~(1)$.
This describes a static compositional modulation of the A2 phase
by a wavevector \textbf{k}$_{0}$=2$\pi$(1,1,1)/$a$
to yield the partially-ordered B2 FeAl (''pseudo-FeAl''), see, e.g.,
Ref. \onlinecite{Khachaturyan_p45}.

{\par} Characterization is performed on samples  at finite
  temperatures with $ 0 < \eta(T) < 1$.
The temperature dependence of $\eta$(T) can only be
obtained from a thermodynamic calculation or measurement.
In Fe-Al, $\eta$(T) varies continuously in the solid phase
(for DO$_{3}$ with arbitrary x, it can be done via two
LRO parameters for the two operative wavevectors, i.e, (111) and 
($\frac{1}{2}\frac{1}{2}\frac{1}{2}$) \cite{Khachaturyan_p45}).
Hence, a Landau expansion of the free energy difference in terms of
F($\eta$) (relative to $\eta $=0 state) for a given fixed-sized unit cell,
DO$_{3}$-type in our case, is written  as
\begin{eqnarray}
\label{Eq.1}
\Delta F^{\sigma}(\eta )
       %&=&  F^{\sigma}(\eta )-F^{\sigma}(\eta =0)
       &=& F^{(2),\sigma}(0)\eta ^{2}+O(\eta ^{4}) \nonumber  \\
        &=& E^{\sigma}(\eta )-E^{\sigma}(0)-T\Delta S(\eta) \, ,
\end{eqnarray}
where odd powers of $\eta$ are zero due to the
translational symmetry of the A2 ($\eta =0$) phase, and
$\Delta$S is an entropy difference for the magnetic state
$\sigma=$(DLM, FM, or NM).
We note that the largest
contribution to the
entropy is the point entropy and this cancels exactly
within a common unit cell; also, when the nearest-neighbor  (chemical)
environments are approximately the same (as roughly is the case here),
the  nearest-neighbor pair entropy cancels exactly.
Therefore, the temperature effects are primarily due to antisite disorder, and
(for fixed composition) can be accounted for by analyzing
total energy differences.
(See  Ref. \onlinecite{JohnsonPRBRC2000} for how such
calculations are related to characterization experiments.)

{\par} In addition to calculations of B2 FeAl versus $\eta$,
fully ordered FeAl and Fe$_{3}$Al, we also consider Fe$_{1-x}$Al$_{x}$
with Al concentration between 25\% and 50\%:
fully-disordered and off-stoichiometric-disordered
B2 and DO$_{3}$ with and without chemical disorder on the Al sublattices
(marked \emph{2} and \emph{4} in Fig. \ref{FigCell}).
Specifically, the following cases of constituent concentrations are
investigated:

(i) For \( 0.25  \leq x\leq 0.50 \): 
\( c_{1}^{Al} =c_{3}^{Al}=0 \), \( c_{4}^{Al}=1 \), \( c_{2}^{Al}=4x-1 \);

(ii) For \( 0.25\leq x\leq 0.50 \): 
\( c_{1}^{Al}=c_{3}^{Al}=0 \), \( c_{2}^{Al}=c_{4}^{Al}=2x \);

(iii) For \( 0.25\leq x\leq 0.25+x_{0}/2 \): 
\( c_{4}^{Al}=1-x_{0} \), \( c_{1}^{Al}=c_{2}^{Al}=c_{3}^{Al}=(4x-(1-x_{0}))/3 \);

(iv) For \( 0.25+x_{0}/2\leq x\leq 0.40 \): 
\( c_{4}^{Al}=1-x_{0} \), \( c_{1}^{Al} \)=\( c_{3}^{Al} \)=x\( _{0} \),
\( c_{2}^{Al} \)=\( (4x-1-x_{0}) \);

(v) For \( 0.25\leq x\leq 0.25+x_{0}/2 \): 
\( c_{1}^{Al}=c_{3}^{Al}=(4x-(1-x_{0}))/3 \), \( c_{2}^{Al}=c_{4}^{Al}=0.5-x_{0}/3 \);

(vi) For \( 0.25+x_{0}/2\leq x\leq 0.50 \): 
\( c_{1}^{Al}=c_{3}^{Al}=x_{0} \), \( c_{2}^{Al}=c_{4}^{Al}=2x-x_{0} \).

Cases (i) and (ii) have no disorder on sublattices \emph{1} and \emph{3}
and \emph{ideal} (no LRO) DO$_{3}$ and B2 compositions
for off-stoichiometric Fe$_{1-x}$Al$_{x}$, (iii)-(vi) are
used to simulate possible compositions with LRO, i.e. thermodynamic
chemical disorder. Concentrations of the constituents for sublattices
\emph{1} and \emph{3} are assumed to be the same both in DO$_{3}$
and B2, and the Al content is not more than
x$_{0}$ (see Fig. \ref{FigCell} caption).
The value of x$_{0}$=0.09 is chosen somewhat arbitrarily for illustrative 
purpose, although it is comparable with the disorder reported 
experimentally from DO$_{3}$-type Fe-Al in Ref.\cite{BradleyJay1932}; 
for B2 FeAl it corresponds $\eta =1-2x_{0}=0.82$.

\section{Paramagnetism in B2 FeAl}
{\par}
As mentioned, experimentally B2 FeAl is found to be paramagnetic,
with finite local moments.
Yet, in agreement with previous theoretical work
\cite{WilliamsKuebler1979,ChachamGalvao1987,SundararajanSahu1995,BognerSteiner1998,KulikovPostnikov1999}
our calculations find that the \emph{ideal}  B2 FeAl is ferromagnetic.
At the theoretical equilibrium lattice parameter $a=5.34$ Bohr, we obtain a magnetic
moment of $\approx 0.72 \mu_{B}$ for Fe and $-0.03 \mu_{B}$ for Al,
again consistent with the published data. However, such calculations
ignore the fact that B2 FeAl is not fully ordered at any finite temperature.

\begin{figure*}
{\centering \resizebox*{3in}{4in}{\includegraphics{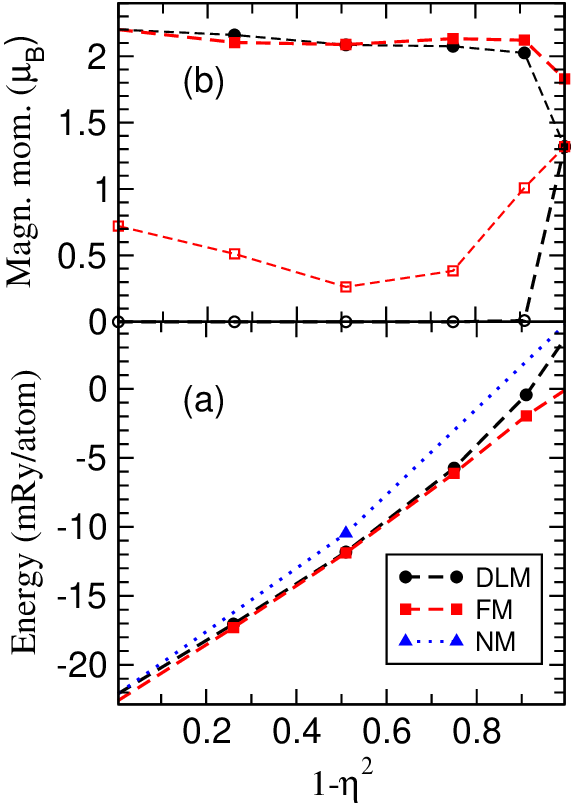}} \par}
%{\centering \resizebox*{3in}{4in}{\includegraphics{FeAlFig2.pdf}} \par}
\caption{The DLM (circles), FM (squares) and NM (triangles)
(a) total energies relative to the A2 FM (mRy/atom) and  (b) Fe magnetic
moments ($\mu_B$) versus partial order $1-\eta ^{2}$, with  filled symbols
  the Al-rich sites (i.e., the Fe anti-sites).}
\label{FigEn5050}
\end{figure*}

{\par}Therefore, in Fig. \ref{FigEn5050}, we show the stability of FM,
NM and DLM partially ordered phases of FeAl alloy versus
long-range order, $\eta$, relative to that of the fully-disordered FM state.

{\par}The NM-FM energy difference decreases monotonically as
 $\eta \rightarrow 1$ (increasing order) from 4.8 mRy/atom in the A2 
 state to 0.5 mRy/atom in the fully ordered B2 structure. 
 For $\eta =0$ (A2) the DLM-FM energy difference, $\Delta E^{DLM-FM}$,
 is $\sim 3.8$ mRy/atom (\( a^{FM}=5.45 \) Bohr) 
with \( \mu ^{FM}_{Fe}=1.83\mu _{B} \) 
 and \( \mu ^{DLM}_{Fe}=1.32\mu _{B} \), respectively. 
We find that the FM state is nominally always the lowest energy state 
for any value of $\eta$
 ($\Delta E^{DLM-FM} \leq 0.25$ mRy/atom, i.e., less than thermal energies,
 for $0.9 \geq \eta \geq 0.5$), except at $\eta \sim 0.7$ ($1-\eta ^{2}=0.5$) 
where the FM state is degenerate with the DLM state. 
For this nearly ordered B2 structure, $\eta =0.7$ corresponds to a 15\% 
concentration of Al on the Fe rich sublattice, compatible with $\sim19\%$ 
from experiment \cite{HumeRothery1969p172,BognerSteiner1998}.
Both the FM and DLM states have nearly the same lattice parameter
($5.36$~Bohr). 
The local Fe anti-sites moments are finite for any partial long-range
order, see Figure \ref{FigEn5050}.  Whereas Fe atoms on Al-rich sites have
nearly the same magnetic moments ($\sim 2.1\mu _{B}$)
for both FM and DLM states,  the Fe moments for the other sites only
exist in the FM state where they are relatively small.
The average PM local moments are comparable to the observed 
Curie-Weiss moments $\sim 0.3 \mu_{B}$ at $\eta = 0.7$.

{\par} It is clear that antisite defects play a crucial role in determining the 
magnetic configuration. However, in the DLM state at $\eta <0.3$ (see
Fig. \ref{FigEn5050}b), the Fe magnetic moments exist in both high-
and low-spin states, similar to bcc Fe; while in the fully-ordered
B2-phase ($\eta =1$), the DLM state is equivalent to the NM state,
where all local moments are equal to zero (quenched). As an aside,
we note that  the PM state (stabilized, as identified here, by  the antisite 
disorder) may benefit from further stabilization afforded by the 
LDA+U procedure\cite{MohnPersson2001,PetukhovMazin2003},
but it is not the origin for the PM effect, as is now clear.

The key point is that the FM and DLM states can be degenerate or nearly
degenerate in the partially ordered equiatomic FeAl state.
Furthermore, Monte Carlo simulations \cite{Restrepo2001} of
A2 FeAl (i.e. $\eta =0$) also obtain a paramagnetic state at room temperature.
The important quantity in understanding non-zero temperature phenomena is the
free energy and the disordered B2-DLM magnetic phase has a larger
configurational entropy contribution to the free energy than the B2-FM
ordered structure over a wide range of $\eta$.
Thus, partial long range chemical order is stable, making the
paramagnetic DLM state the stable configuration at finite temperature, 
even with only a small degree of disorder.

{\par} We note that the theoretical DLM lattice constant is about 3\%
below the experimental value ($\approx 5.50$ Bohr). However, such a deviation
is comparable with pure bcc-Fe, as found in other calculations.
The calculated DLM magnetic moment on an antisite Fe atom
(i.e. on Al-rich sites) is close to the value of 2.2 $\mu _{B}$
  observed by Parthasarathi and Beck \cite{ParthasarathiBeck1976}
  for samples quenched in cold water. In addition, Bogner \emph{et al.}\cite{BognerSteiner1998} have extracted a moment of
  $\approx 3.43 \mu _{B}$ from Langevin-type behavior of
magnetization-field dependence.   Bogner \emph{et al.} hypothesized that
such a moment could result from a cluster formed by nine nearest
Fe atoms, which would correspond to a partially-ordered alloy with
a concentration of Fe atoms on Al sites of $\sim  18\%$,  close to the 19.12\% 
found in experiment. Again, it is similar to the disorder we find  needed
for  degeneracy in our calculations, i.e.,  $\sim15\%$  with $\eta = 1-2(0.15) = 0.7$.

\section{Structural Anomalies in Fe-rich Fe-Al}
\begin{figure}
{\centering \resizebox*{3in}{3.66in}{\includegraphics{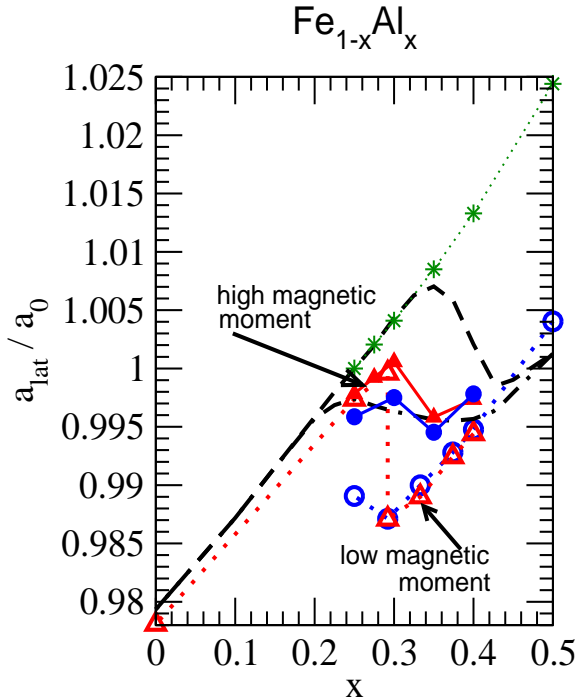}} \par}
%{\centering \resizebox*{3in}{3in}{\includegraphics{FeAlFig3.pdf}} \par}
\caption{Reduced lattice parameter of Fe$_{1-x}$Al$_{x}$ versus x.
Theory results are for \emph{off-stoichiometric} DO$_{3}$ (triangles)
and B2 (circles), or the A2 phases (stars). Theoretical
$a_{0}  (5.320$ Bohr) is the lattice constant of the A2 at x=0.25.
Filled symbols (triangles/circles) are for DO$_{3}$/B2  with
additional thermal (antisite) disorder, see text.
Experimental data is the dashed (dash-dotted) line for
as-deformed samples (quenched from $1273$~K)
\protect\cite{HumeRothery1969p172}, and $a_{0}$ is that observed
lattice constant at x=0.25 for deformed sample.
Quenching from $523$~K yields similar results to high-T quench data.}
  \label{FigAvsX}
\end{figure}

{\par}
Lattice constant anomalies have been experimentally observed in the
Fe-rich region of the Fe-Al phase diagram. The lattice constant dependence
on concentration is strongly dependent on the quenching or annealing
procedure \cite{HumeRothery1969p172} used in synthesizing the material
(dashed line in Fig. \ref{FigAvsX}), i.e. thermal antisites and vacancies.
For low Al concentration the lattice constant has a nearly linear dependence
on concentration. However, between 25\%-40\%
the lattice parameter exhibits non-monotonic behavior with
a maximum lattice constant at \( x<0.3 \) (for deformed samples
at $x\sim 0.34$) and a minimum between $0.34<x<0.43$ with increasing
Al content. Of course, the exact behavior strongly depends
on the type of quenching procedure used during processing
\cite{HumeRothery1969p172}.
On the other hand, the experimental data do not provide quantitative information
on (chemical and magnetic) sample-dependent disorder in the
observed \cite{BradleyJay1932} partially ordered DO$_{3}$ and B2 structures,
which have chemical disorder on each independent sublattice (estimated
to about $8\%$ near $x=0.25$) and partial long-range order.
It is assumed that the formation of the ''pseudo-FeAl'' type of ordering
is associated with the structural anomalies \cite{HumeRothery1969p172}.

First, to address the anomalies, we consider the ''ideal'' DO$_{3}$:
case (i) -- with disorder only on the Al sublattice \emph{4}; and
ideal  B2: case (ii) -- with disorder on the Al sublattices \emph{2} and \emph{4}.

{\par}Our DO$_{3}$ calculations exhibit linear growth of the lattice
constant with increasing concentration up to $x\sim 0.29$, while
the average magnetic moment (per atom) $\mu$ decreases from
$1.42 \mu _{B}$ at $x=0.25$ to $1.23 \mu _{B}$. At \( x\simeq 0.29 \)
two degenerate solutions exist for the FM state. The \emph{low-spin}
solution has both a smaller $\mu \sim 0.54 \mu _{B}$ and a smaller lattice
constant (see Fig. \ref{FigAvsX}). However, the largest discrepancy
is for the Fe magnetic moments on sites \emph{1} and \emph{3} where
the \emph{low-spin} solution has a significantly smaller
magnetic moment on these sites ($\sim 0.2\mu_{B}$) as compared
to \emph{high-spin} solution ($\sim 1.52\mu_{B}$)
on the same sites. Above x=0.29, the \emph{low-spin} solution
is energetically more favorable; at elevated temperatures a
smooth transition from one solution to the second solution should be expected
due to the possible coexistence of both solutions in a range of Al concentration.

\begin{figure}
{\centering \resizebox*{3in}{3in}{\includegraphics{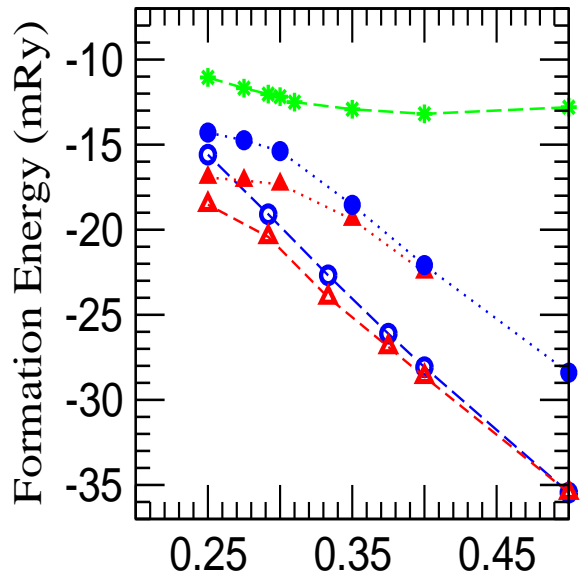}} \par}
%{\centering \resizebox*{3in}{3in}{\includegraphics{FeAlFig4.pdf}} \par}
\caption{Formation energy of ferromagnetic A2, DO$_{3}$
and B2 Fe$_{1-x}$Al$_{x}$ versus concentration of Al; symbols
are chosen as in Fig. \ref{FigAvsX}.}
\label{FigEvsX}
\end{figure}

{\par} In addition we note that the calculated lattice parameter of the
fully-disordered A2 phase is almost linear for concentrations of Al
up to 34\% which agrees well with experimental observations
  for deformed samples quenched at high temperatures,
see Fig. \ref{FigAvsX}. For the A2 phase the energy is about $7$ mRy
larger than ''ideal'' DO$_{3}$ at $x=0.25$, see Fig. \ref{FigEvsX}.
The respective energy difference begins to grow rapidly with increasing
Al concentration at $x\sim 0.3$: from $8.4$ mRy at $x=0.29$ to
$22.6$ mRy at $x=0.5$.

{\par}Consideration of the ''ideal'' B2-disordered Fe$_{1-x}$A$_{x}$
reveals that the lattice parameter is practically indistinguishable
from the \emph{low-spin} solution in DO$_{3}$ calculations,
however, with magnetic moments that are
slightly higher than the ones found in DO$_{3}$.
The difference in the total energy between \emph{pseudo}-FeAl
and ''ideal'' DO$_{3}$ declines rapidly, see Fig. \ref{FigEvsX}.
from 2.9 mRy/atom at x=0.25 to 0.5 mRy/atom at x=0.40 and finally to zero
at x=0.5, where the chemical structures are equivalent.
Taking into account that \emph{pseudo}-FeAl configurational
entropy contribution to free energy is larger than
in partially-ordered DO$_{3}$ for the same composition, it is
reasonable to expect that at finite temperature off-stoichiometric
DO$_{3}$ and B2 phases could coexist for some values of x.
For  high enough Al content DO$_{3}$ would eventually be
replaced by the B2 phase as the ground state where again the actual
value for the Al concentration would depend on both the temperature
and processing procedure.

{\par}To illustrate the importance of the thermodynamical chemical disorder
we performed calculations for both DO$_{3}$ (iii,iv) and B2 (v,vi) phases.
Figure \ref{FigAvsX} shows a stronger lattice constant dependence on
concentration for B2 than for the "high-spin'' DO$_{3}$ solution
(i.e. at $x<0.3$). For larger Al content the influence on B2 and DO$_{3}$
is comparable. In addition, the linear dependence
of the formation energy on the Al concentration is virtually unaffected
by the chemical disorder (excluding the uniform shift), see Fig. \ref{FigEvsX},
for $x>0.34$. However, for smaller Al concentrations this is not the case.
This is due to the  dependence of the sublattice composition on Al content
in (iii)-(vi). The additional effect of partial-order on top of off-stoichiometric
disorder leads roughly to a plateau in the formation energy, especially
for DO$_3$
(see Fig.  \ref{FigEvsX}), which is compatible with the shift off stoichiometry
for the maximum in the B2-DO$_3$ transition temperature.

{\par} Generally, the magneto-chemical coupling effect is a ubiquitous 
phenomena observed in many materials. It has been used also  to
explain quantitatively, e.g., ordering characterization data
\cite{JohnsonPRBRC2000} and INVAR lattice anomaly in Fe-Ni
\cite{CrisanEntel2002}. 

\section{Summary}
{\par}
The structural and magnetic anomalies observed in bcc-based Fe-Al have
remained unexplained experimentally and theoretically for over 50 years.
As characterization data and their interpretation depend upon the sample
preparation and processing, the analysis of their properties based on
simulation require the inclusion of these important thermal effects.
We presented a theoretical study that  incorporates
magnetic, off-stoichiometric and thermodynamic chemical disorder
on equal footing with the electronic structure, exemplifying
a quantitative means for studying materials processed at 
high temperatures where point defects (such as anti-sites) 
play an important role in determining the properties.
We find that partial long-range order due to thermal (i.e. antisite) disorder
yields paramagnetic B2 FeAl (with finite moments) that is stable with respect 
to the ferromagnetic state,  in contrast to past theory but
 in agreement with experiment. The partial order required also
agrees with experiment.
In addition, we showed that this same magneto-chemical coupling 
produces structural anomalies observed versus Al content.
Our results provide a clear and simple explanation behind the
observed paramagnetism and structural anomalies in Fe-Al.
We expect that this type of long-range order effect is responsible for
the observed Fe$_3$Al lattice constant dependence on annealing 
temperature\cite{Selisskiy1957}, as well as the paramagnetism
(or spin-glass behavior\cite{ShuklaWortis1980}) in B2-Fe$_{1-x}$Al$_{x}$,
both of which may be investigated using the approach employed here
for Fe$_{0.5}$Al$_{0.5}$.

\begin{acknowledgments}
This research is supported by the Mathematical, Information and Computational
Sciences, Office of Advanced Scientific Computing Research, U.S. Department of
Energy at Oak Ridge National Laboratory under Contract DE-AC05-00OR22725
with UT-Battelle Limited Liability Corporation, and at the Frederick Seitz
Materials Research Laboratory under contract DEFG02-91ER45439.
%Accordingly, the U.S. Government retains a non-exclusive, royalty-free license
%to publish or reproduce the published form of this contribution, or
%allow others to do so, for U.S. Government purposes.
\end{acknowledgments}

%\newpage

\bibliographystyle{apsrev}
\bibliography{FeAl}

\begin{thebibliography}{44}
\expandafter\ifx\csname natexlab\endcsname\relax\def\natexlab#1{#1}\fi
\expandafter\ifx\csname bibnamefont\endcsname\relax
  \def\bibnamefont#1{#1}\fi
\expandafter\ifx\csname bibfnamefont\endcsname\relax
  \def\bibfnamefont#1{#1}\fi
\expandafter\ifx\csname citenamefont\endcsname\relax
  \def\citenamefont#1{#1}\fi
\expandafter\ifx\csname url\endcsname\relax
  \def\url#1{\texttt{#1}}\fi
\expandafter\ifx\csname urlprefix\endcsname\relax\def\urlprefix{URL }\fi
\providecommand{\bibinfo}[2]{#2}
\providecommand{\eprint}[2][]{\url{#2}}

\bibitem[{\citenamefont{Besnus et~al.}(1975)\citenamefont{Besnus, Herr, and
  Meyer}}]{BesnusHerr1975}
\bibinfo{author}{\bibfnamefont{M.~J.} \bibnamefont{Besnus}},
  \bibinfo{author}{\bibfnamefont{A.}~\bibnamefont{Herr}}, \bibnamefont{and}
  \bibinfo{author}{\bibfnamefont{A.~J.~P.} \bibnamefont{Meyer}},
  \bibinfo{journal}{J. of Phys. F} \textbf{\bibinfo{volume}{5}},
  \bibinfo{pages}{2138} (\bibinfo{year}{1975}).

\bibitem[{\citenamefont{Parthasarathi and Beck}(1976)}]{ParthasarathiBeck1976}
\bibinfo{author}{\bibfnamefont{A.}~\bibnamefont{Parthasarathi}}
  \bibnamefont{and} \bibinfo{author}{\bibfnamefont{P.~A.} \bibnamefont{Beck}},
  \bibinfo{journal}{Solid State Commun.} \textbf{\bibinfo{volume}{18}},
  \bibinfo{pages}{211} (\bibinfo{year}{1976}).

\bibitem[{\citenamefont{Domke and Thomas}(1984)}]{DomkeThomas1984}
\bibinfo{author}{\bibfnamefont{H.}~\bibnamefont{Domke}} \bibnamefont{and}
  \bibinfo{author}{\bibfnamefont{L.~K.} \bibnamefont{Thomas}},
  \bibinfo{journal}{J. of Magn. Magn. Mater.} \textbf{\bibinfo{volume}{45}},
  \bibinfo{pages}{305} (\bibinfo{year}{1984}).

\bibitem[{\citenamefont{Bogner et~al.}(1998)\citenamefont{Bogner, Steiner,
  Reissner, Mohn, Blaha, Schwarz, Krachler, Ipser, and
  Sepiol}}]{BognerSteiner1998}
\bibinfo{author}{\bibfnamefont{J.}~\bibnamefont{Bogner}},
  \bibinfo{author}{\bibfnamefont{W.}~\bibnamefont{Steiner}},
  \bibinfo{author}{\bibfnamefont{M.}~\bibnamefont{Reissner}},
  \bibinfo{author}{\bibfnamefont{P.}~\bibnamefont{Mohn}},
  \bibinfo{author}{\bibfnamefont{P.}~\bibnamefont{Blaha}},
  \bibinfo{author}{\bibfnamefont{K.}~\bibnamefont{Schwarz}},
  \bibinfo{author}{\bibfnamefont{R.}~\bibnamefont{Krachler}},
  \bibinfo{author}{\bibfnamefont{H.}~\bibnamefont{Ipser}}, \bibnamefont{and}
  \bibinfo{author}{\bibfnamefont{B.}~\bibnamefont{Sepiol}},
  \bibinfo{journal}{Phys. Rev. B} \textbf{\bibinfo{volume}{58}},
  \bibinfo{pages}{14922} (\bibinfo{year}{1998}).

\bibitem[{\citenamefont{Williams et~al.}(1979)\citenamefont{Williams,
  K{\"u}bler, and C.~D.~Gellat}}]{WilliamsKuebler1979}
\bibinfo{author}{\bibfnamefont{A.~R.} \bibnamefont{Williams}},
  \bibinfo{author}{\bibfnamefont{J.}~\bibnamefont{K{\"u}bler}},
  \bibnamefont{and}
  \bibinfo{author}{\bibfnamefont{J.}~\bibnamefont{C.~D.~Gellat}},
  \bibinfo{journal}{Phys. Rev. B} \textbf{\bibinfo{volume}{19}},
  \bibinfo{pages}{6094} (\bibinfo{year}{1979}).

\bibitem[{\citenamefont{Chacham et~al.}(1987)\citenamefont{Chacham, da~Silva,
  Guenzburger, and Ellis}}]{ChachamGalvao1987}
\bibinfo{author}{\bibfnamefont{H.}~\bibnamefont{Chacham}},
  \bibinfo{author}{\bibfnamefont{E.~G.} \bibnamefont{da~Silva}},
  \bibinfo{author}{\bibfnamefont{D.}~\bibnamefont{Guenzburger}},
  \bibnamefont{and} \bibinfo{author}{\bibfnamefont{D.~E.} \bibnamefont{Ellis}},
  \bibinfo{journal}{Phys. Rev. B} \textbf{\bibinfo{volume}{35}},
  \bibinfo{pages}{1602} (\bibinfo{year}{1987}).

\bibitem[{\citenamefont{Sundararajan et~al.}(1995)\citenamefont{Sundararajan,
  Sahu, Kanhere, Panat, and Das}}]{SundararajanSahu1995}
\bibinfo{author}{\bibfnamefont{V.}~\bibnamefont{Sundararajan}},
  \bibinfo{author}{\bibfnamefont{B.~R.} \bibnamefont{Sahu}},
  \bibinfo{author}{\bibfnamefont{D.~G.} \bibnamefont{Kanhere}},
  \bibinfo{author}{\bibfnamefont{P.~V.} \bibnamefont{Panat}}, \bibnamefont{and}
  \bibinfo{author}{\bibfnamefont{G.~P.} \bibnamefont{Das}},
  \bibinfo{journal}{J. of Phys.: Condens. Matter} \textbf{\bibinfo{volume}{7}},
  \bibinfo{pages}{6019} (\bibinfo{year}{1995}).

\bibitem[{\citenamefont{Bose et~al.}(1997)\citenamefont{Bose, Drchal,
  Kudrnovsky, Jepsen, and Andersen}}]{Bose1997}
\bibinfo{author}{\bibfnamefont{S.~K.} \bibnamefont{Bose}},
  \bibinfo{author}{\bibfnamefont{V.}~\bibnamefont{Drchal}},
  \bibinfo{author}{\bibfnamefont{J.}~\bibnamefont{Kudrnovsky}},
  \bibinfo{author}{\bibfnamefont{O.}~\bibnamefont{Jepsen}}, \bibnamefont{and}
  \bibinfo{author}{\bibfnamefont{O.~K.} \bibnamefont{Andersen}},
  \bibinfo{journal}{Phys. Rev. B} \textbf{\bibinfo{volume}{55}},
  \bibinfo{pages}{8184} (\bibinfo{year}{1997}).

\bibitem[{\citenamefont{Kulikov et~al.}(1999)\citenamefont{Kulikov, Postnikov,
  Borstel, and Braun}}]{KulikovPostnikov1999}
\bibinfo{author}{\bibfnamefont{N.~I.} \bibnamefont{Kulikov}},
  \bibinfo{author}{\bibfnamefont{A.~V.} \bibnamefont{Postnikov}},
  \bibinfo{author}{\bibfnamefont{G.}~\bibnamefont{Borstel}}, \bibnamefont{and}
  \bibinfo{author}{\bibfnamefont{J.}~\bibnamefont{Braun}},
  \bibinfo{journal}{Phys. Rev. B} \textbf{\bibinfo{volume}{59}},
  \bibinfo{pages}{6824} (\bibinfo{year}{1999}).

\bibitem[{\citenamefont{Hume-Rothery et~al.}(1969)\citenamefont{Hume-Rothery,
  Smallman, and Haworth}}]{HumeRothery1969p172}
\bibinfo{author}{\bibfnamefont{W.}~\bibnamefont{Hume-Rothery}},
  \bibinfo{author}{\bibfnamefont{R.~E.} \bibnamefont{Smallman}},
  \bibnamefont{and} \bibinfo{author}{\bibfnamefont{C.~W.}
  \bibnamefont{Haworth}}, \emph{\bibinfo{title}{The Structure of Metals and
  Alloys}} (\bibinfo{publisher}{The Metals and Metallurgy Trust of the
  Institute of Metals and Institution of Metallurgists},
  \bibinfo{address}{London}, \bibinfo{year}{1969}), pp.
  \bibinfo{pages}{172--173}.

\bibitem[{\citenamefont{Baker and Munroe}(1990)}]{BakerMunroe1990}
\bibinfo{author}{\bibfnamefont{I.}~\bibnamefont{Baker}} \bibnamefont{and}
  \bibinfo{author}{\bibfnamefont{P.~R.} \bibnamefont{Munroe}}, in
  \emph{\bibinfo{booktitle}{High Temperature Aluminides and Intermetallics}},
  edited by \bibinfo{editor}{\bibfnamefont{S.~H.} \bibnamefont{Whang}},
  \bibinfo{editor}{\bibfnamefont{C.}~\bibnamefont{Liu}},
  \bibinfo{editor}{\bibfnamefont{D.}~\bibnamefont{Pope}}, \bibnamefont{and}
  \bibinfo{editor}{\bibfnamefont{J.}~\bibnamefont{Stiegler}}
  (\bibinfo{publisher}{The Minerals, Metals and Material Society},
  \bibinfo{year}{1990}), pp. \bibinfo{pages}{425--452}.

\bibitem[{\citenamefont{Liu et~al.}(1989)\citenamefont{Liu, Lee, and
  Mckamey}}]{LiuLee1989}
\bibinfo{author}{\bibfnamefont{C.~T.} \bibnamefont{Liu}},
  \bibinfo{author}{\bibfnamefont{E.~H.} \bibnamefont{Lee}}, \bibnamefont{and}
  \bibinfo{author}{\bibfnamefont{C.~G.} \bibnamefont{Mckamey}},
  \bibinfo{journal}{Scr. Metall.} \textbf{\bibinfo{volume}{23}},
  \bibinfo{pages}{875} (\bibinfo{year}{1989}).

\bibitem[{\citenamefont{Takahashi and
  Y.Umakoshi}(1991)}]{TakahashiUmakoshi1991}
\bibinfo{author}{\bibfnamefont{S.}~\bibnamefont{Takahashi}} \bibnamefont{and}
  \bibinfo{author}{\bibnamefont{Y.Umakoshi}}, \bibinfo{journal}{J. Phys.:
  Condens. Matter} \textbf{\bibinfo{volume}{3}}, \bibinfo{pages}{5805}
  (\bibinfo{year}{1991}).

\bibitem[{\citenamefont{Nagpal and Baker}(1990)}]{NagpalBaker1990}
\bibinfo{author}{\bibfnamefont{P.}~\bibnamefont{Nagpal}} \bibnamefont{and}
  \bibinfo{author}{\bibfnamefont{I.}~\bibnamefont{Baker}},
  \bibinfo{journal}{Metall. Trans.} \textbf{\bibinfo{volume}{21A}},
  \bibinfo{pages}{2281} (\bibinfo{year}{1990}).

\bibitem[{\citenamefont{Chang et~al.}(1993)\citenamefont{Chang, Pike, Liu,
  Bilbrey, and Stone}}]{ChangPike1993}
\bibinfo{author}{\bibfnamefont{Y.~A.} \bibnamefont{Chang}},
  \bibinfo{author}{\bibfnamefont{L.~M.} \bibnamefont{Pike}},
  \bibinfo{author}{\bibfnamefont{C.~T.} \bibnamefont{Liu}},
  \bibinfo{author}{\bibfnamefont{A.~R.} \bibnamefont{Bilbrey}},
  \bibnamefont{and} \bibinfo{author}{\bibfnamefont{D.~S.} \bibnamefont{Stone}},
  \bibinfo{journal}{J. Intermetall.} \textbf{\bibinfo{volume}{1}},
  \bibinfo{pages}{107} (\bibinfo{year}{1993}).

\bibitem[{\citenamefont{Weber et~al.}(1997)\citenamefont{Weber, Meurtin, Paris,
  Fourdeux, and Lesbats}}]{WeberMeurtin1997}
\bibinfo{author}{\bibfnamefont{D.}~\bibnamefont{Weber}},
  \bibinfo{author}{\bibfnamefont{M.}~\bibnamefont{Meurtin}},
  \bibinfo{author}{\bibfnamefont{D.}~\bibnamefont{Paris}},
  \bibinfo{author}{\bibfnamefont{A.}~\bibnamefont{Fourdeux}}, \bibnamefont{and}
  \bibinfo{author}{\bibfnamefont{P.}~\bibnamefont{Lesbats}},
  \bibinfo{journal}{J. Phys. C} \textbf{\bibinfo{volume}{7}},
  \bibinfo{pages}{332} (\bibinfo{year}{1997}).

\bibitem[{\citenamefont{Fu et~al.}(1993)\citenamefont{Fu, Ye, Yoo, and
  Ho}}]{FuYe1993}
\bibinfo{author}{\bibfnamefont{C.~L.} \bibnamefont{Fu}},
  \bibinfo{author}{\bibfnamefont{Y.~Y.} \bibnamefont{Ye}},
  \bibinfo{author}{\bibfnamefont{M.~H.} \bibnamefont{Yoo}}, \bibnamefont{and}
  \bibinfo{author}{\bibfnamefont{K.~M.} \bibnamefont{Ho}},
  \bibinfo{journal}{Phys. Rev. B} \textbf{\bibinfo{volume}{48}},
  \bibinfo{pages}{6712} (\bibinfo{year}{1993}).

\bibitem[{\citenamefont{Lundqvist and March}(1983)}]{DFTReviewLundqvist1983}
\bibinfo{editor}{\bibfnamefont{S.}~\bibnamefont{Lundqvist}} \bibnamefont{and}
  \bibinfo{editor}{\bibfnamefont{N.~H.} \bibnamefont{March}}, eds.,
  \emph{\bibinfo{title}{Theory of the Inhomogeneous Gas}}
  (\bibinfo{publisher}{Plenum}, \bibinfo{address}{New York},
  \bibinfo{year}{1983}).

\bibitem[{\citenamefont{Els{\"a}sser et~al.}(1998)\citenamefont{Els{\"a}sser,
  Zhu, Louie, F{\"a}ahnle, and Chan}}]{ElsaesserZhu1998}
\bibinfo{author}{\bibfnamefont{C.}~\bibnamefont{Els{\"a}sser}},
  \bibinfo{author}{\bibfnamefont{J.}~\bibnamefont{Zhu}},
  \bibinfo{author}{\bibfnamefont{S.~G.} \bibnamefont{Louie}},
  \bibinfo{author}{\bibfnamefont{M.}~\bibnamefont{F{\"a}ahnle}},
  \bibnamefont{and} \bibinfo{author}{\bibfnamefont{C.~T.} \bibnamefont{Chan}},
  \bibinfo{journal}{J. Phys.: Condens. Matter} \textbf{\bibinfo{volume}{10}},
  \bibinfo{pages}{5081} (\bibinfo{year}{1998}).

\bibitem[{\citenamefont{Lechermann et~al.}(2002)\citenamefont{Lechermann,
  Welsch, Els{\"a}sser, Ederer, and F{\"a}hnle}}]{LechermannWelsch2002}
\bibinfo{author}{\bibfnamefont{F.}~\bibnamefont{Lechermann}},
  \bibinfo{author}{\bibfnamefont{F.}~\bibnamefont{Welsch}},
  \bibinfo{author}{\bibfnamefont{C.}~\bibnamefont{Els{\"a}sser}},
  \bibinfo{author}{\bibfnamefont{C.}~\bibnamefont{Ederer}}, \bibnamefont{and}
  \bibinfo{author}{\bibfnamefont{M.}~\bibnamefont{F{\"a}hnle}},
  \bibinfo{journal}{Phys. Rev. B} \textbf{\bibinfo{volume}{65}},
  \bibinfo{pages}{132104} (\bibinfo{year}{2002}).

\bibitem[{\citenamefont{Mohn et~al.}(2001)\citenamefont{Mohn, Persson, Blaha,
  Schwarz, Novak, and Eschrig}}]{MohnPersson2001}
\bibinfo{author}{\bibfnamefont{P.}~\bibnamefont{Mohn}},
  \bibinfo{author}{\bibfnamefont{C.}~\bibnamefont{Persson}},
  \bibinfo{author}{\bibfnamefont{P.}~\bibnamefont{Blaha}},
  \bibinfo{author}{\bibfnamefont{K.}~\bibnamefont{Schwarz}},
  \bibinfo{author}{\bibfnamefont{P.}~\bibnamefont{Novak}}, \bibnamefont{and}
  \bibinfo{author}{\bibfnamefont{H.}~\bibnamefont{Eschrig}},
  \bibinfo{journal}{Phys. Rev. Lett.} \textbf{\bibinfo{volume}{87}},
  \bibinfo{pages}{196401} (\bibinfo{year}{2001}).

\bibitem[{\citenamefont{Petukhov et~al.}(2003)\citenamefont{Petukhov, Mazin,
  Chioncel, and Lichtenstein}}]{PetukhovMazin2003}
\bibinfo{author}{\bibfnamefont{A.~G.} \bibnamefont{Petukhov}},
  \bibinfo{author}{\bibfnamefont{I.~I.} \bibnamefont{Mazin}},
  \bibinfo{author}{\bibfnamefont{L.}~\bibnamefont{Chioncel}}, \bibnamefont{and}
  \bibinfo{author}{\bibfnamefont{A.~I.} \bibnamefont{Lichtenstein}},
  \bibinfo{journal}{Phys. Rev. B} \textbf{\bibinfo{volume}{67}},
  \bibinfo{pages}{153106} (\bibinfo{year}{2003}).

\bibitem[{\citenamefont{Bradley and Jay}(1932)}]{BradleyJay1932}
\bibinfo{author}{\bibfnamefont{A.~J.} \bibnamefont{Bradley}} \bibnamefont{and}
  \bibinfo{author}{\bibfnamefont{A.~H.} \bibnamefont{Jay}},
  \bibinfo{journal}{Proc. R. Soc. London} \textbf{\bibinfo{volume}{136}},
  \bibinfo{pages}{210} (\bibinfo{year}{1932}).

\bibitem[{\citenamefont{Taylor and Jones}(1958)}]{TaylorJones1958}
\bibinfo{author}{\bibfnamefont{A.}~\bibnamefont{Taylor}} \bibnamefont{and}
  \bibinfo{author}{\bibfnamefont{R.~M.} \bibnamefont{Jones}},
  \bibinfo{journal}{Phys. Chem. Solids} \textbf{\bibinfo{volume}{6}},
  \bibinfo{pages}{16} (\bibinfo{year}{1958}).

\bibitem[{\citenamefont{Lawley and Cahn}(1961)}]{LawleyCahn1961}
\bibinfo{author}{\bibfnamefont{A.}~\bibnamefont{Lawley}} \bibnamefont{and}
  \bibinfo{author}{\bibfnamefont{R.~W.} \bibnamefont{Cahn}},
  \bibinfo{journal}{Phys. Chem. Solids} \textbf{\bibinfo{volume}{20}},
  \bibinfo{pages}{204} (\bibinfo{year}{1961}).

\bibitem[{\citenamefont{Johnson et~al.}(2000)\citenamefont{Johnson, Smirnov,
  Staunton, Pinski, and Shelton}}]{JohnsonPRBRC2000}
\bibinfo{author}{\bibfnamefont{D.~D.} \bibnamefont{Johnson}},
  \bibinfo{author}{\bibfnamefont{A.~V.} \bibnamefont{Smirnov}},
  \bibinfo{author}{\bibfnamefont{J.~B.} \bibnamefont{Staunton}},
  \bibinfo{author}{\bibfnamefont{F.~J.} \bibnamefont{Pinski}},
  \bibnamefont{and} \bibinfo{author}{\bibfnamefont{W.~A.}
  \bibnamefont{Shelton}}, \bibinfo{journal}{Phys. Rev. B}
  \textbf{\bibinfo{volume}{62}}, \bibinfo{pages}{R11917}
  (\bibinfo{year}{2000}).

\bibitem[{\citenamefont{Korringa}(1947)}]{Korringa1947}
\bibinfo{author}{\bibfnamefont{J.}~\bibnamefont{Korringa}},
  \bibinfo{journal}{Physica} \textbf{\bibinfo{volume}{13}},
  \bibinfo{pages}{392} (\bibinfo{year}{1947}).

\bibitem[{\citenamefont{Kohn and Rostoker}(1954)}]{KohnRostoker1954}
\bibinfo{author}{\bibfnamefont{W.}~\bibnamefont{Kohn}} \bibnamefont{and}
  \bibinfo{author}{\bibfnamefont{N.}~\bibnamefont{Rostoker}},
  \bibinfo{journal}{Phys. Rev.} \textbf{\bibinfo{volume}{94}},
  \bibinfo{pages}{1111} (\bibinfo{year}{1954}).

\bibitem[{\citenamefont{von Barth and Hedin}(1972)}]{BarthJPhysC1972}
\bibinfo{author}{\bibfnamefont{U.}~\bibnamefont{von Barth}} \bibnamefont{and}
  \bibinfo{author}{\bibfnamefont{L.}~\bibnamefont{Hedin}}, \bibinfo{journal}{J.
  Phys. C} \textbf{\bibinfo{volume}{5}}, \bibinfo{pages}{1629}
  (\bibinfo{year}{1972}).

\bibitem[{Cal()}]{CalcDetails}
\bibinfo{note}{All angular momentum expansions include up to l$_{max}$=3; a
  semi-circular contour in complex plane with 18 points is used to integrate
  the Green's function over energy; at each energy, a Brillouin zone
  integration is performed by a special k-points method with 144 k-points. Such
  choices provides better than $0.1 mRy/atom$ relative accuracy in total
  energy.}

\bibitem[{\citenamefont{Johnson et~al.}(1986)\citenamefont{Johnson, Nicholson,
  Pinski, Gy${\ddot o}$rffy, and Stocks}}]{Johnsonetal1986}
\bibinfo{author}{\bibfnamefont{D.~D.} \bibnamefont{Johnson}},
  \bibinfo{author}{\bibfnamefont{D.~M.} \bibnamefont{Nicholson}},
  \bibinfo{author}{\bibfnamefont{F.~J.} \bibnamefont{Pinski}},
  \bibinfo{author}{\bibfnamefont{B.~L.} \bibnamefont{Gy${\ddot o}$rffy}},
  \bibnamefont{and} \bibinfo{author}{\bibfnamefont{G.~M.}
  \bibnamefont{Stocks}}, \bibinfo{journal}{Phys. Rev. Lett.}
  \textbf{\bibinfo{volume}{56}}, \bibinfo{pages}{2088} (\bibinfo{year}{1986}).

\bibitem[{\citenamefont{Johnson et~al.}(1990)\citenamefont{Johnson, Nicholson,
  Pinski, Gy${\ddot o}$rffy, and Stocks}}]{Johnsonetal1990}
\bibinfo{author}{\bibfnamefont{D.~D.} \bibnamefont{Johnson}},
  \bibinfo{author}{\bibfnamefont{D.~M.} \bibnamefont{Nicholson}},
  \bibinfo{author}{\bibfnamefont{F.~J.} \bibnamefont{Pinski}},
  \bibinfo{author}{\bibfnamefont{B.~L.} \bibnamefont{Gy${\ddot o}$rffy}},
  \bibnamefont{and} \bibinfo{author}{\bibfnamefont{G.~M.}
  \bibnamefont{Stocks}}, \bibinfo{journal}{Phys. Rev. B}
  \textbf{\bibinfo{volume}{41}}, \bibinfo{pages}{9701} (\bibinfo{year}{1990}).

\bibitem[{\citenamefont{Johnson and Pinski}(1993)}]{JohnsonPinski1993}
\bibinfo{author}{\bibfnamefont{D.~D.} \bibnamefont{Johnson}} \bibnamefont{and}
  \bibinfo{author}{\bibfnamefont{F.~J.} \bibnamefont{Pinski}},
  \bibinfo{journal}{Phys. Rev. B} \textbf{\bibinfo{volume}{48}},
  \bibinfo{pages}{11553} (\bibinfo{year}{1993}).

\bibitem[{\citenamefont{Hubbard}(1983)}]{Hubbard1983}
\bibinfo{author}{\bibfnamefont{J.}~\bibnamefont{Hubbard}},
  \bibinfo{journal}{Phys. Rev. Lett.} \textbf{\bibinfo{volume}{51}},
  \bibinfo{pages}{300} (\bibinfo{year}{1983}).

\bibitem[{\citenamefont{Hasagawa}(1979)}]{Hasegawa1979}
\bibinfo{author}{\bibfnamefont{H.}~\bibnamefont{Hasagawa}},
  \bibinfo{journal}{J. Phys. Soc. Jpn.} \textbf{\bibinfo{volume}{46}},
  \bibinfo{pages}{1504} (\bibinfo{year}{1979}).

\bibitem[{\citenamefont{Staunton et~al.}(1985)\citenamefont{Staunton, Gy${\ddot
  o}$rffy, Pindor, Stocks, and Winter}}]{Staunton1985}
\bibinfo{author}{\bibfnamefont{J.~B.} \bibnamefont{Staunton}},
  \bibinfo{author}{\bibfnamefont{B.}~\bibnamefont{Gy${\ddot o}$rffy}},
  \bibinfo{author}{\bibfnamefont{A.~J.} \bibnamefont{Pindor}},
  \bibinfo{author}{\bibfnamefont{G.~M.} \bibnamefont{Stocks}},
  \bibnamefont{and} \bibinfo{author}{\bibfnamefont{H.}~\bibnamefont{Winter}},
  \bibinfo{journal}{J. Phys. F: Met. Phys.} \textbf{\bibinfo{volume}{15}},
  \bibinfo{pages}{1387} (\bibinfo{year}{1985}).

\bibitem[{\citenamefont{Johnson et~al.}(1987)\citenamefont{Johnson, Pinski, and
  Staunton}}]{Johnson1987}
\bibinfo{author}{\bibfnamefont{D.~D.} \bibnamefont{Johnson}},
  \bibinfo{author}{\bibfnamefont{F.~J.} \bibnamefont{Pinski}},
  \bibnamefont{and} \bibinfo{author}{\bibfnamefont{J.~B.}
  \bibnamefont{Staunton}}, \bibinfo{journal}{J. Appl. Phys.}
  \textbf{\bibinfo{volume}{61}}, \bibinfo{pages}{3715} (\bibinfo{year}{1987}).

\bibitem[{\citenamefont{Johnson et~al.}(1989)\citenamefont{Johnson, Pinski,
  Staunton, Gy${\ddot o}$rffy, and Stocks}}]{Johnsonetal1989}
\bibinfo{author}{\bibfnamefont{D.~D.} \bibnamefont{Johnson}},
  \bibinfo{author}{\bibfnamefont{F.~J.} \bibnamefont{Pinski}},
  \bibinfo{author}{\bibfnamefont{J.~B.} \bibnamefont{Staunton}},
  \bibinfo{author}{\bibfnamefont{B.~L.} \bibnamefont{Gy${\ddot o}$rffy}},
  \bibnamefont{and} \bibinfo{author}{\bibfnamefont{G.~M.}
  \bibnamefont{Stocks}}, in \emph{\bibinfo{booktitle}{Physical Metallurgy of
  Controlled Expansion "INVAR-type" Alloys}}, edited by
  \bibinfo{editor}{\bibfnamefont{K.}~\bibnamefont{Russell}} \bibnamefont{and}
  \bibinfo{editor}{\bibfnamefont{D.}~\bibnamefont{Smith}}
  (\bibinfo{publisher}{Minerals, Metals, Materials Society},
  \bibinfo{address}{Materials Park, OH}, \bibinfo{year}{1989}), pp.
  \bibinfo{pages}{3--24}.

\bibitem[{\citenamefont{Johnson and Shelton}(1997)}]{JohnsonShelton1997}
\bibinfo{author}{\bibfnamefont{D.~D.} \bibnamefont{Johnson}} \bibnamefont{and}
  \bibinfo{author}{\bibfnamefont{W.~A.} \bibnamefont{Shelton}}, in
  \emph{\bibinfo{booktitle}{The INVAR Effect - A Centennial Symposium}}, edited
  by \bibinfo{editor}{\bibfnamefont{J.}~\bibnamefont{Wittenauer}}
  (\bibinfo{publisher}{Minerals, Metals, Materials Society},
  \bibinfo{address}{Materials Park, OH}, \bibinfo{year}{1997}), p.
  \bibinfo{pages}{6374}.

\bibitem[{\citenamefont{Khachaturyan}(1983)}]{Khachaturyan_p45}
\bibinfo{author}{\bibfnamefont{A.~G.} \bibnamefont{Khachaturyan}},
  \emph{\bibinfo{title}{Theory of Structural Phase Transformations in Solids}}
  (\bibinfo{publisher}{Wiley}, \bibinfo{address}{New York},
  \bibinfo{year}{1983}), pp. \bibinfo{pages}{45--46}.

\bibitem[{\citenamefont{Restrepo et~al.}(2001)\citenamefont{Restrepo,
  Alc{\'a}zar, and Landau}}]{Restrepo2001}
\bibinfo{author}{\bibfnamefont{J.}~\bibnamefont{Restrepo}},
  \bibinfo{author}{\bibfnamefont{G.~A.~P.} \bibnamefont{Alc{\'a}zar}},
  \bibnamefont{and} \bibinfo{author}{\bibfnamefont{D.~P.}
  \bibnamefont{Landau}}, \bibinfo{journal}{J. Appl. Phys.}
  \textbf{\bibinfo{volume}{89}}, \bibinfo{pages}{7341} (\bibinfo{year}{2001}).

\bibitem[{\citenamefont{Crisan et~al.}(2002)\citenamefont{Crisan, Entel, Ebert,
  Akai, Johnson, and Staunton}}]{CrisanEntel2002}
\bibinfo{author}{\bibfnamefont{V.}~\bibnamefont{Crisan}},
  \bibinfo{author}{\bibfnamefont{P.}~\bibnamefont{Entel}},
  \bibinfo{author}{\bibfnamefont{H.}~\bibnamefont{Ebert}},
  \bibinfo{author}{\bibfnamefont{H.}~\bibnamefont{Akai}},
  \bibinfo{author}{\bibfnamefont{D.~D.} \bibnamefont{Johnson}},
  \bibnamefont{and} \bibinfo{author}{\bibfnamefont{J.~B.}
  \bibnamefont{Staunton}}, \bibinfo{journal}{Phys. Rev. B}
  \textbf{\bibinfo{volume}{66}}, \bibinfo{pages}{014416}
  (\bibinfo{year}{2002}).

\bibitem[{\citenamefont{Selisskiy}(1957)}]{Selisskiy1957}
\bibinfo{author}{\bibfnamefont{Y.~P.} \bibnamefont{Selisskiy}},
  \bibinfo{journal}{Fiz. Metal. i Metalloved.} \textbf{\bibinfo{volume}{4}},
  \bibinfo{pages}{191} (\bibinfo{year}{1957}).

\bibitem[{\citenamefont{Shukla and Wortis}(1980)}]{ShuklaWortis1980}
\bibinfo{author}{\bibfnamefont{P.}~\bibnamefont{Shukla}} \bibnamefont{and}
  \bibinfo{author}{\bibfnamefont{M.}~\bibnamefont{Wortis}},
  \bibinfo{journal}{Phys. Rev. B} \textbf{\bibinfo{volume}{21}},
  \bibinfo{pages}{159} (\bibinfo{year}{1980}).

\end{thebibliography}

\end{document}